\documentclass{aastex701}

\newcommand{\vsini}{\mbox{$v \sin i$}}

\begin{document}

\shorttitle{Doppler imaging of UX~Ari}
\shortauthors{Xiang et al.}

\title{Starspot activity and surface differential rotation on UX~Arietis}

\affiliation{International Centre of Supernovae (ICESUN), Yunnan Key Laboratory of Supernova Research, Yunnan Observatories, Chinese Academy of Sciences, Kunming 650216, China}
\affiliation{Yunnan Observatories, Chinese Academy of Sciences, Kunming 650216, China}

\author{Yue Xiang}
\affiliation{International Centre of Supernovae (ICESUN), Yunnan Key Laboratory of Supernova Research, Yunnan Observatories, Chinese Academy of Sciences, Kunming 650216, China}
\affiliation{Key Laboratory for the Structure and Evolution of Celestial Objects, Chinese Academy of Sciences, Kunming 650216, China}
\email[show]{xy@ynao.ac.cn}

\author{Shenghong Gu}
\affiliation{Yunnan Observatories, Chinese Academy of Sciences, Kunming 650216, China}
\affiliation{Key Laboratory for the Structure and Evolution of Celestial Objects, Chinese Academy of Sciences, Kunming 650216, China}
\affiliation{School of Astronomy and Space Science, University of Chinese Academy of Sciences, Beijing 101408, China}
\email[show]{shenghonggu@ynao.ac.cn}

\author{A. Collier Cameron}
\affiliation{School of Physics and Astronomy, University of St Andrews, Fife KY16 9SS, UK}
\email{}

\author{J. R. Barnes}
\affiliation{Department of Physical Sciences, The Open University, Walton Hall, Milton Keynes MK7 6AA, UK}
\email{}

\author{Dongtao Cao}
\affiliation{International Centre of Supernovae (ICESUN), Yunnan Key Laboratory of Supernova Research, Yunnan Observatories, Chinese Academy of Sciences, Kunming 650216, China}
\affiliation{Key Laboratory for the Structure and Evolution of Celestial Objects, Chinese Academy of Sciences, Kunming 650216, China}
\email{}

\correspondingauthor{Yue Xiang, Shenghong Gu}

\begin{abstract}

We present new Doppler images of the K0 subgiant primary component of the RS CVn-type binary UX~Arietis (UX~Ari), derived from time-series spectra obtained in November--December of 2017 and 2024. Observations demonstrate that some spectral lines of the K0~IV component exhibit rapid changes on timescales of 1-2 hours, which seem not to be resulting from spot activity, meanwhile other spectral lines show no such fast variations. Through an investigation, we find that the Ca I 6439 \AA\ profile shows variation that follows the rotational modulation of spots. Using this line as a reference, we derive the least-squares deconvolution (LSD) profile from the selected lines of each spectrum so as to generate a more reliable Doppler image, which is consistent with the shape of the corresponding Ca I 6439 \AA\ line. The Doppler images are separately reconstructed from the Ca I 6439 \AA\ and the LSD profiles for each dataset, and the surface maps are in good agreement with each other. All of the surface maps show dominant starspot structure at mid-to-high latitudes with appendages extending to the equator, while their locations differ by about 0.5 in the rotational phase between 2017 and 2024. In 2017 November-December, the main starspot group appears to be spatially associated with a large flare event just half a month later. Through the cross-correlation method, we have derived a weak anti-solar differential rotation for the primary component of UX~Ari, while its equator belt is well tidally locked.

\end{abstract}


\section{Introduction}

Starspots are the manifestations of stellar magnetic field. Tracking of starspots allows us not only to indirectly investigate stellar magnetic structures but also to measure rotations of stars, a key factor of the stellar dynamo process. The Doppler imaging technique converts time-series high-resolution spectral lines into surface starspot maps \citep{vogt1987}, which can provide detailed information on the starspot distribution and evolution and also enable the measurements of surface differential rotation rates for rapidly rotating stars \citep{petit2002,barnes2017,kovari2024}. 

UX~Arietis (UX Ari, HD 21242) is a double-lined RS CVn-type binary, consisting of a K0 subgiant primary and a G5 main-sequence secondary components, with an orbital period of about 6.44 days \citep{carlos1971}. The stellar age of UX~Ari is 5.6 Gyr, and its two components have masses of 1.30 and 1.14 $M_{\odot}$, radii of 5.6 and 1.6 $R_{\odot}$, respectively \citep{hummel2017}. The spectra of UX~Ari indicate the presence of an additional star, which is not a member of the system but coincidentally lies on the same line of sight \citep{duemmler2001}. Another unseen component of UX~Ari was revealed by \citet{duemmler2001} through analyzing variations in the systemic radial velocity of the inner binary. Using radio astrometry, \citet{peterson2011} inferred a mass of $0.75 M_{\odot}$ for this tertiary component.

The primary component of UX~Ari is magnetically active. \citet{donati1992} detected significant Zeeman signatures in the Stokes V spectra of UX~Ari, which were attributed to the primary component. UX~Ari shows strong chromospheric activity, such as high-rate flares \citep{carlos1971,frasca1994,gu2002,ekmekci2010,cao2017}. A prominent eruption event was reported by \citet{cao2025}, which was associated with a huge flare located on the opposite hemisphere of the primary component in 2017 December. UX~Ari also shows significant brightness variations caused by starspots on surface of the primary component \citep{aarum2003,rosario2007}. Interestingly, UX~Ari becomes bluer when it is fainter, which may be explained by the presence of bright faculae \citep{aarum2003b}. The G5~V component of UX~Ari is also active, but has a much lower level than the primary one \citep{messina2008}.

Doppler imaging studies were conducted for UX~Ari by several authors \citep{vogt1991,aarum1999,gu2004}, which revealed a wide distribution of starspots on the surface of K subgiant component. By tracing the evolution of individual starspots, \citet{vogt1991} found a weak anti-solar differential rotation for the primary component of UX~Ari. Its pole rotates $1.09\degr {\rm d}^{-1}$ ($0.019~{\rm rad~d^{-1}}$) faster than the equator.

To further investigate the starspot distribution and evolution as well as measure the surface differential rotation rate, we apply Doppler imaging to UX~Ari based on time-series spectra collected in November--December of 2017 and 2024. In section 2, we describe the spectroscopic observations and data reduction. In section 3, we present the new Doppler images and the measurements of differential rotation. Finally, we discuss and summarize the results in section 4.  

\section{Observations and data reduction}

The spectroscopic observations of UX~Ari were carried out in 2017 and 2024, and a total of 89 spectra were collected. From 2017 November to December, the fibre-fed high-resolution spectrograph (HRS) mounted on the 2.16m telescope \citep{fan2016} at the Xinglong station of National Astronomical Observatories of China was used for the observations. In 2024 November--December, besides the 2.16m telescope, we also employed the 2.4m telescope \citep{fan2015} at the Lijiang station of Yunnan observatories. A spectrograph similar to the HRS was used to observe UX~Ari. Due to the different fibre diameter, the spectral resolution of the HRS on the 2.16m telescope is R$\sim$48000 but that on the 2.4m telescope is R$\sim$32000. 4K$\times$4K CCD cameras were used to record data during the observations. The wavelength coverage of both spectrographs is 3900-10000\AA.

The raw image data collected by the two telescopes are reduced by using the IRAF\footnote{IRAF is distributed by the National Optical Astronomy Observatory, which is operated by the Association of Universities for Research in Astronomy (AURA) under cooperative agreement with the National Science Foundation.} package in the standard procedures, including image trimming, bias and flat field corrections, scatter light subtraction, cosmic-ray removal, 1D spectrum extraction, wavelength calibration and continuum fitting. The observations are summarized in Table \ref{tab:log}, including the UTC date, exposure time, Heliocentric Julian Date (HJD), orbital phase, and signal-to-noise ratio (SNR). The orbital phases are calculated by using the ephemeris ${\rm HJD} = 2456238.134 + 6.437888P$ derived by \citet{hummel2017}. The full version of Table \ref{tab:log} is only available in the machine-readable format.

\begin{deluxetable}{lcccccccc}
\tabletypesize{\scriptsize}
\tablecolumns{4}
\tablewidth{0pt}
\tablecaption{A summary of the observations on UX~Ari.}
 \label{tab:log}
\tablehead{
 \colhead{UTC Date}&
 \colhead{Exposure time}&
  \colhead{HJD}&
 \colhead{Phase}&
 \colhead{SNR (6439 \AA)}&
 \colhead{SNR (LSD)}\\
 \colhead{YYYY-MM-DD}&
 \colhead{s}&
 \colhead{2400000+}&
 \colhead{}&
 \colhead{}&
 \colhead{}
}

\startdata
  Xinglong 2.16m\\
  2017-11-28 & 900 & 58086.18749 & 0.059 &  108 & 644\\
  2017-11-28 & 900 & 58086.21450 & 0.063 &   99 & 645\\
  2017-11-28 & 900 & 58086.31627 & 0.079 &  104 & 660\\
  2017-11-29 & 900 & 58087.19964 & 0.216 &  127 & 677\\
  2017-11-29 & 900 & 58087.22587 & 0.220 &  142 & 681\\
  2017-11-29 & 900 & 58087.25160 & 0.224 &  147 & 684\\
\enddata
\tablecomments{Table 1 is published in its entirety in the machine-readable format.}
\end{deluxetable}

\begin{figure}
\centering
\includegraphics[width=0.45\textwidth]{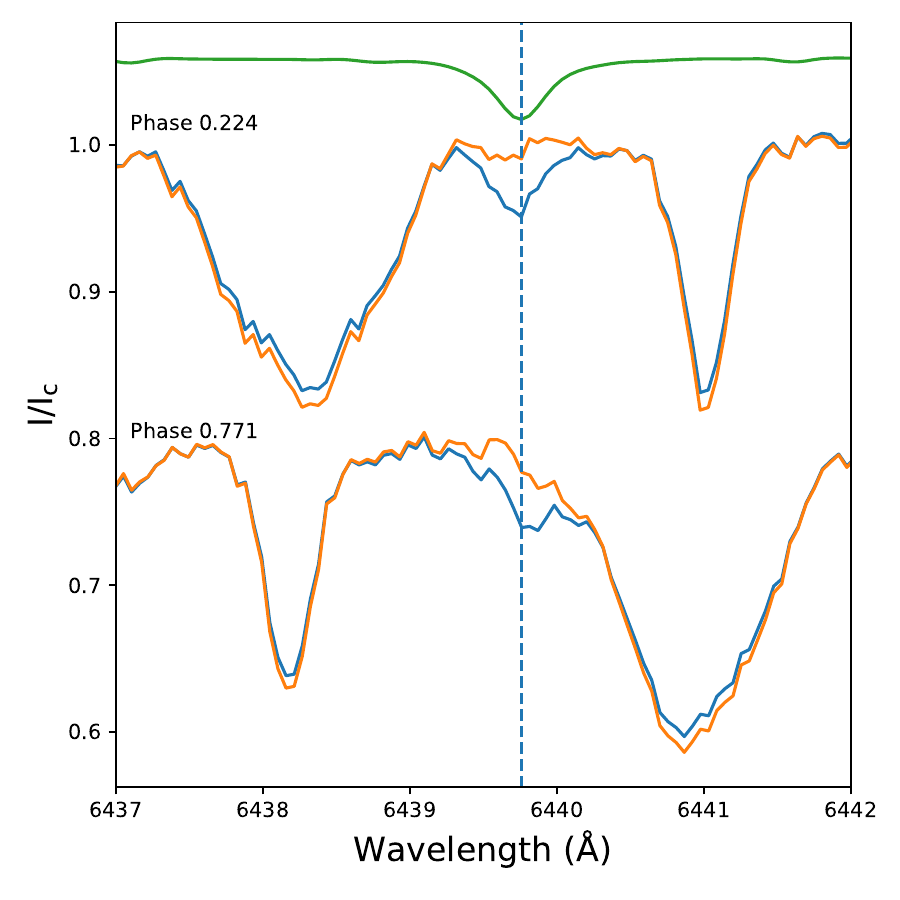}
\caption{Removals of the spectral line of the additional star. The observed spectral profiles are in blue and the corrected ones are in orange, and the removed profile is shown in green. The vertical dashed line marks the position of the additional line.}
\label{fig:ts}
\end{figure}

An additional star's signal is present in the spectrum of UX~Ari. Although it is not part of the UX~Ari system, it lies along the same line of sight. The weak contribution of the additional star to the observed spectra should be removed before Doppler imaging. We have used the spectrum of a K3 main-sequence star, HR 753, as the template, according to the spectral type of the additional star determined by \citet{aarum2003b}. The template is rotationally broadened, shifted in radial velocity and scaled in intensity to fit the additional line of the observed spectral profile, the result is then subtracted from the spectra of UX~Ari. We show examples of such removals in Figure \ref{fig:ts}.

\begin{figure*}
\centering
\includegraphics[width=0.2\textwidth]{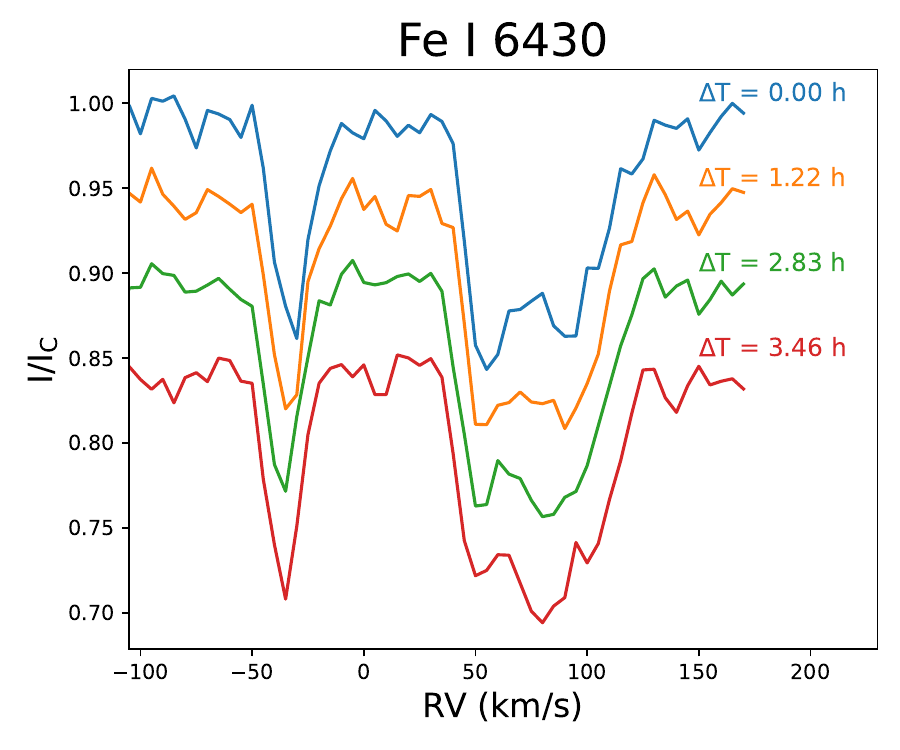}
\includegraphics[width=0.2\textwidth]{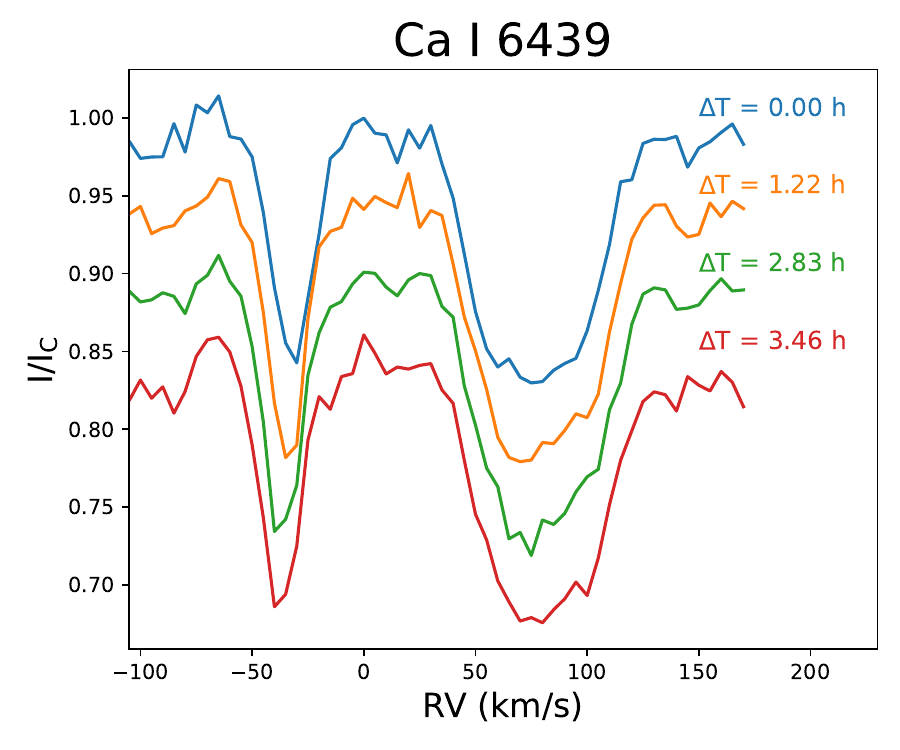}
\includegraphics[width=0.2\textwidth]{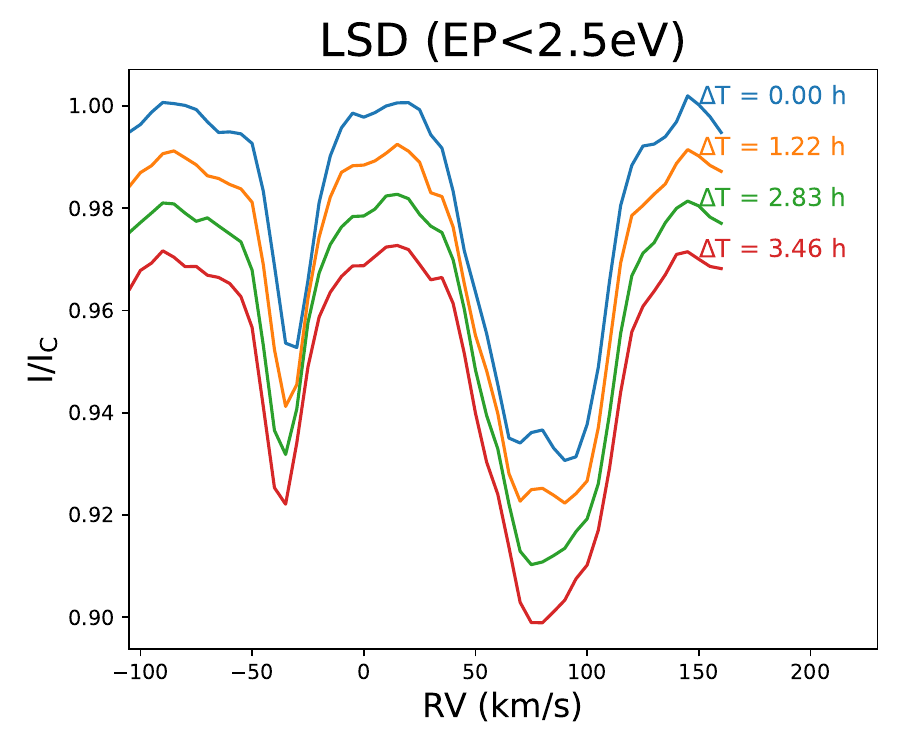}
\includegraphics[width=0.2\textwidth]{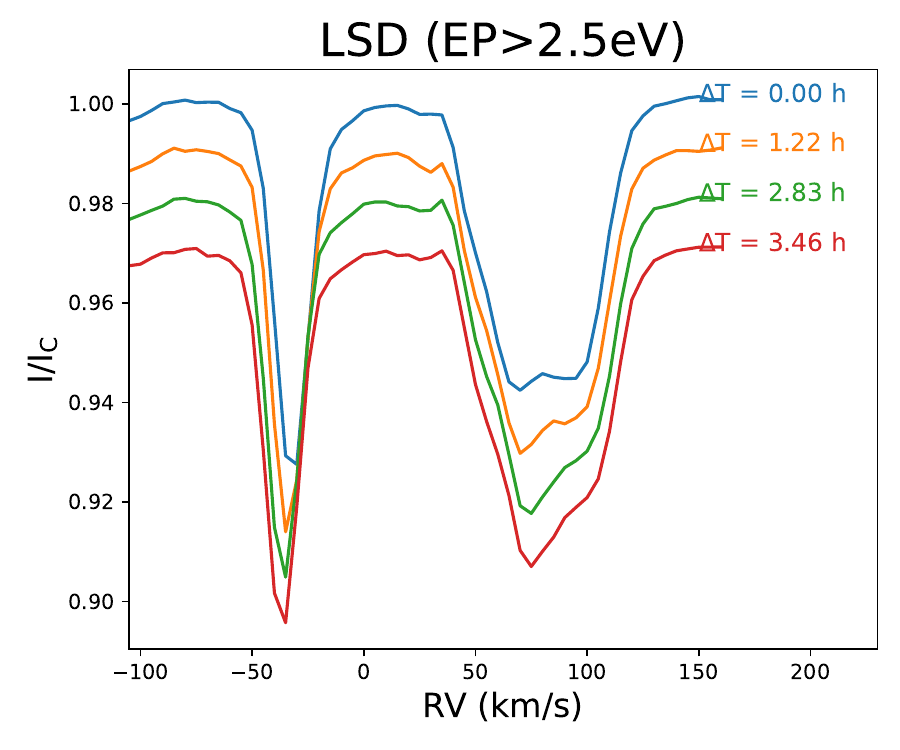}

\includegraphics[width=0.25\textwidth]{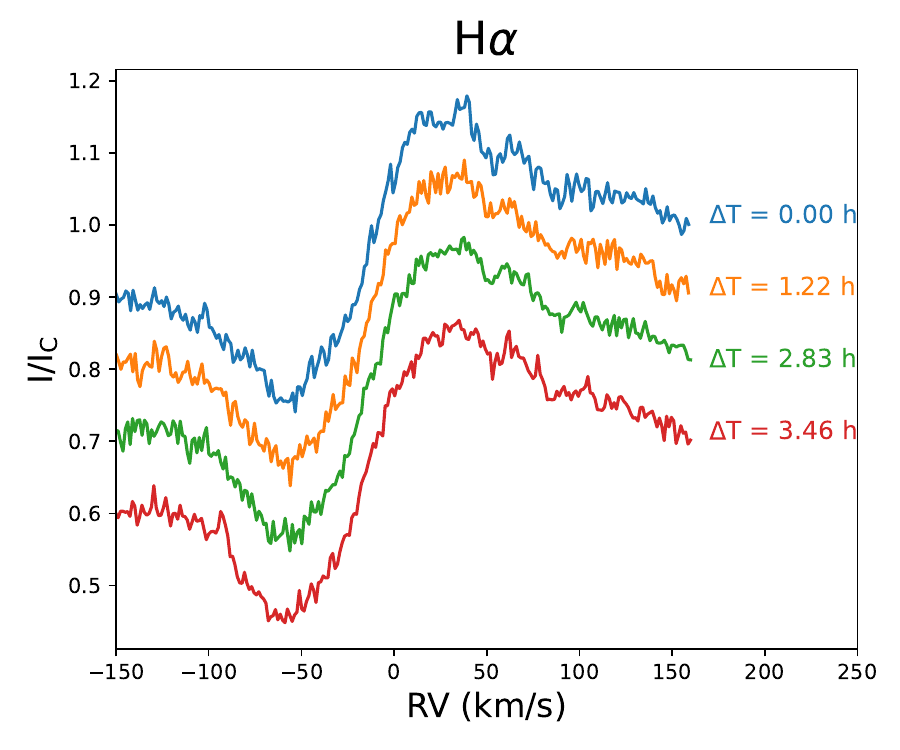}
\includegraphics[width=0.45\textwidth]{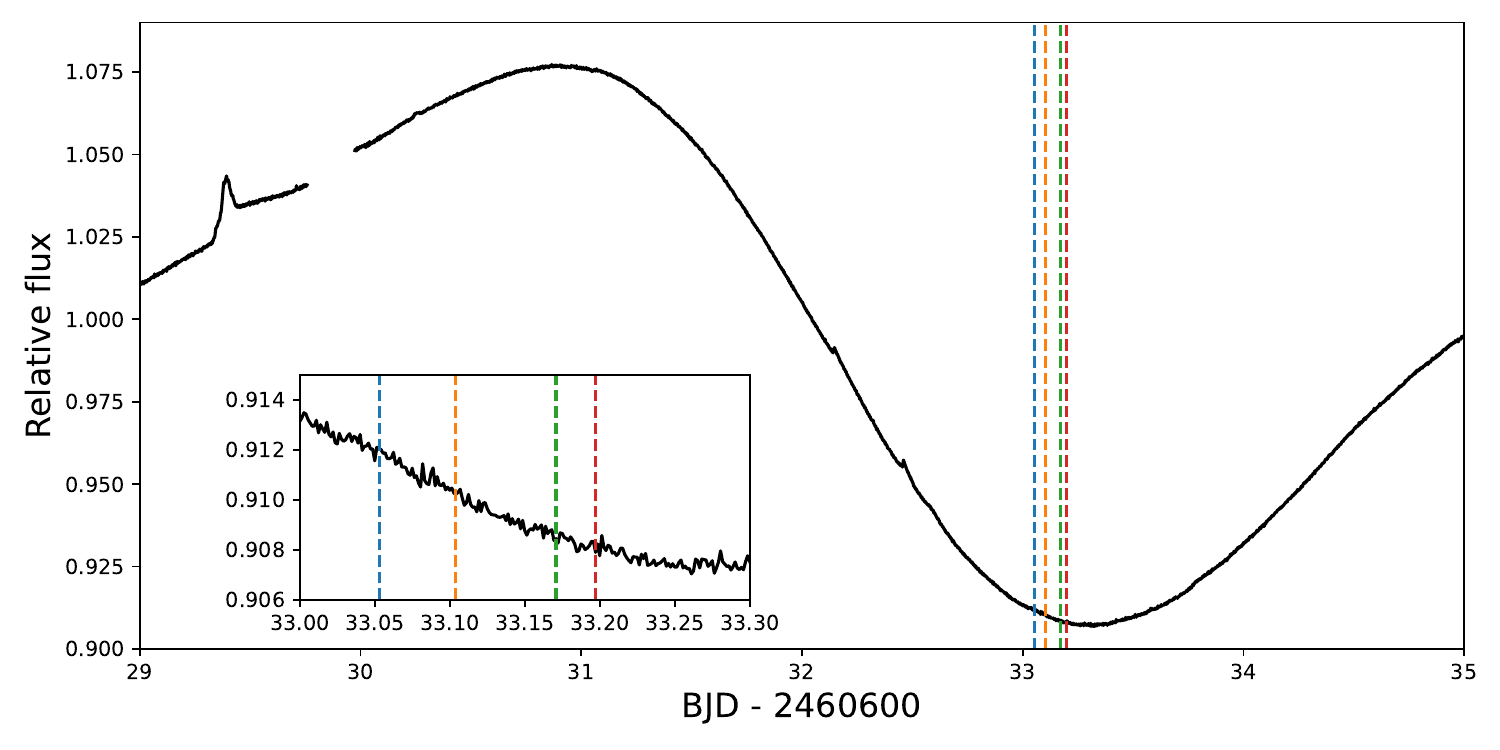}
\caption{Top panels show the Fe I 6430\AA, Ca I 6439\AA\ profiles and two LSD profiles derived from lines with different EPs, observed on 2024 November 18. Bottom panels show the H$\alpha$ lines and the TESS light curve, where the vertical dashed lines mark the corresponding times of four spectroscopic observations in that night.}
\label{fig:example}
\end{figure*}

In general, the least-squares deconvolution (LSD; \citealt{donati1997b}) technique is applied to derive a mean profile with much higher SNR for Doppler imaging. The LSD method is based on the simplified assumption that starspots affect all photospheric lines in a similar way so that all lines share a similar shape. However, UX~Ari appears to be an exception to this assumption. Some of its spectral lines show rapid shape variations, while others do not. An example is shown in Figure \ref{fig:example}. The Fe I 6430 \AA\ and Ca I 6439 \AA\ lines are two commonly used spectral lines for Doppler imaging of late-type stars. In this figure, the Fe I 6430 \AA\ line shows rapid changes within 1-2 hours in one night, whereas the Ca I 6439 \AA\ line exhibits gradual variations consistent with those caused by starspots. In order to get the LSD profiles with higher SNRs, which show the same behaviour as the Ca I 6439 \AA\ profile, we have extracted the line list for UX~Ari from the Vienna Atomic Line Database (VALD; \citealt{kupka1999,ryabchikova2015}) and selected lines with different atomic parameters for tests. We find that the LSD profiles derived from the list of about 900 lines with the excitation potential (EP) larger than 2.5 eV and the oscillator strength ($\log gf$) larger than -1.5 show shape variations consistent with those of the Ca I 6439 \AA\ line, and the SNR improves by a factor of about six. Some LSD profiles derived from the lines with different EPs are also shown in Figure \ref{fig:example}. The rapid variations of the Fe I 6430 \AA\ line and the LSD profile from the lines with lower EPs seem not to be attributed to flaring activity. As shown in Figure \ref{fig:example}, the H$\alpha$ line did not change obviously during that time, and the TESS light curve also showed no rapid variations beyond normal rotational modulation. The lack of the H$\alpha$ and white-light flares implies that the Fe I 6430 \AA\ line perturbations are not due to a chromospheric eruption activity. In the next section, we shall perform Doppler imaging with the Ca I 6439 \AA\ lines and the LSD profiles derived above.

\section{Doppler imaging}

\subsection{Stellar parameters}

The orbital parameters of UX~Ari, such as the radial velocity semiamplitudes of two components ($K_{1}$ and $K_{2}$), the inclination ($i$), the conjunction time (T$_{0}$), and the orbital period (P) are taken from \citet{hummel2017}. Using these orbital parameters, we fine-tune the projected rotation speed (\vsini) for two components. We have performed a fixed number of iterations with the DoTS (Doppler Tomography of Stars) Doppler imaging code \citep{cameron1994,cameron1997} by using a set of parameters and searched for the ones leading to the minimum $\chi^{2}$. This method can suppress the effect of the starspots on the stellar parameters \citep{barnes2005}. We find $K_{1}$ is not satisfied our observations well and derive a better value of 61.3 km/s with the $\chi^{2}$ minimization method. This value is between those derived by \citet{hummel2017} and \citet{duemmler2001}, and in good agreement to that of \citet{massarotti2008}. The final adopted stellar parameters for Doppler imaging are listed in Table \ref{tab:par}.

\begin{deluxetable}{lcc}
\tabletypesize{\scriptsize}
\tablecolumns{3}
\tablewidth{0pt}
\tablecaption{Adopted stellar parameters of UX~Ari for Doppler imaging.}
 \label{tab:par}
\tablehead{
 \colhead{Parameter}& 
 \colhead{Value}&
 \colhead{Ref.}\\
}
\startdata
 $q=M_{2}/M_{1}$ & 0.912 & a\\
 $K_{1}$ (km s$^{-1}$) & 61.3 & DoTS\\
 $K_{2}$ (km s$^{-1}$) & 67.2  & a\\
$i$ (\degr) & 55 & a\\
 T$_{0}$ (HJD) & 2456238.134 & a\\
 P$_{\rm orb}$ (d)  & 6.437888 & a\\
 \vsini~$_{1}$ (km s$^{-1}$) & 36.6 & DoTS\\
 \vsini~$_{2}$ (km s$^{-1}$) & 10.2 & DoTS\\
\enddata
\tablecomments{a: \citet{hummel2017}.}
\end{deluxetable}

\subsection{Surface images}

To reconstruct the surface images of UX~Ari, we have applied the DoTS code in the binary mode, which takes into account the orbital motion and simultaneously reconstructs both surfaces of two components (e.g. \citealt{binary}). The DoTS performs a maximum entropy regularized reconstruction based on a two-temperature model, which assumes the stellar surface is composed of only two components, the hotter photosphere and cooler starspot, and thus produces the surface image representing by the spot filling factor. During the preparation work, we have constructed the look-up tables, which contains local intensities of the photosphere and starspot at different limb angles, by using the spectra of HR~3351 (K0~IV) and HR~3309 (G5~V) for the photospheres of the primary and secondary components of UX~Ari, the spectrum of HR 248 (M0~III) for the starspots, respectively. These template spectra were observed using the same instrument setup as UX Ari. The linear limb darkening coefficients derived by \citet{claret2012} are applied in the generation of the look-up tables.

\begin{figure*}
\gridline{\fig{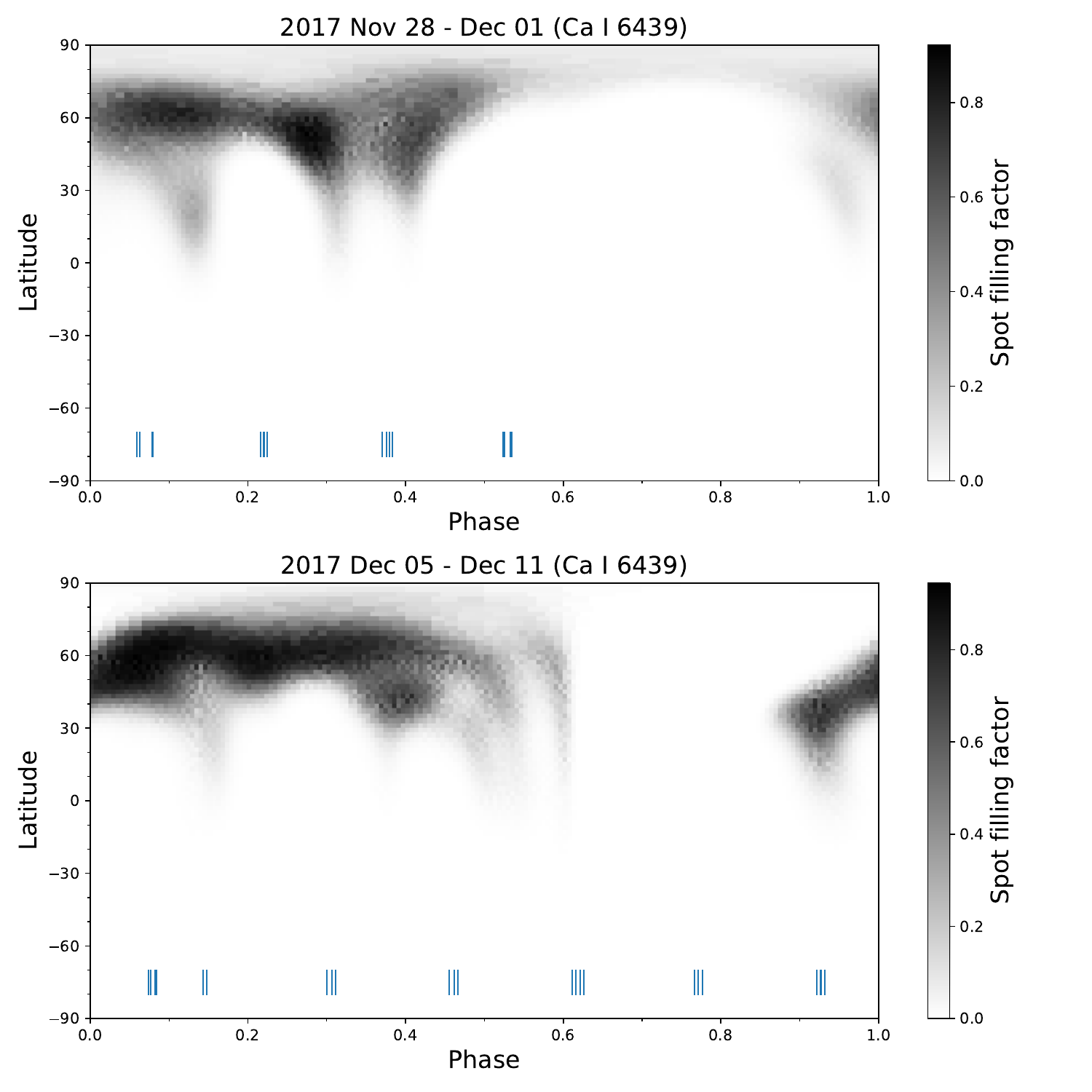}{0.45\textwidth}{}
          \fig{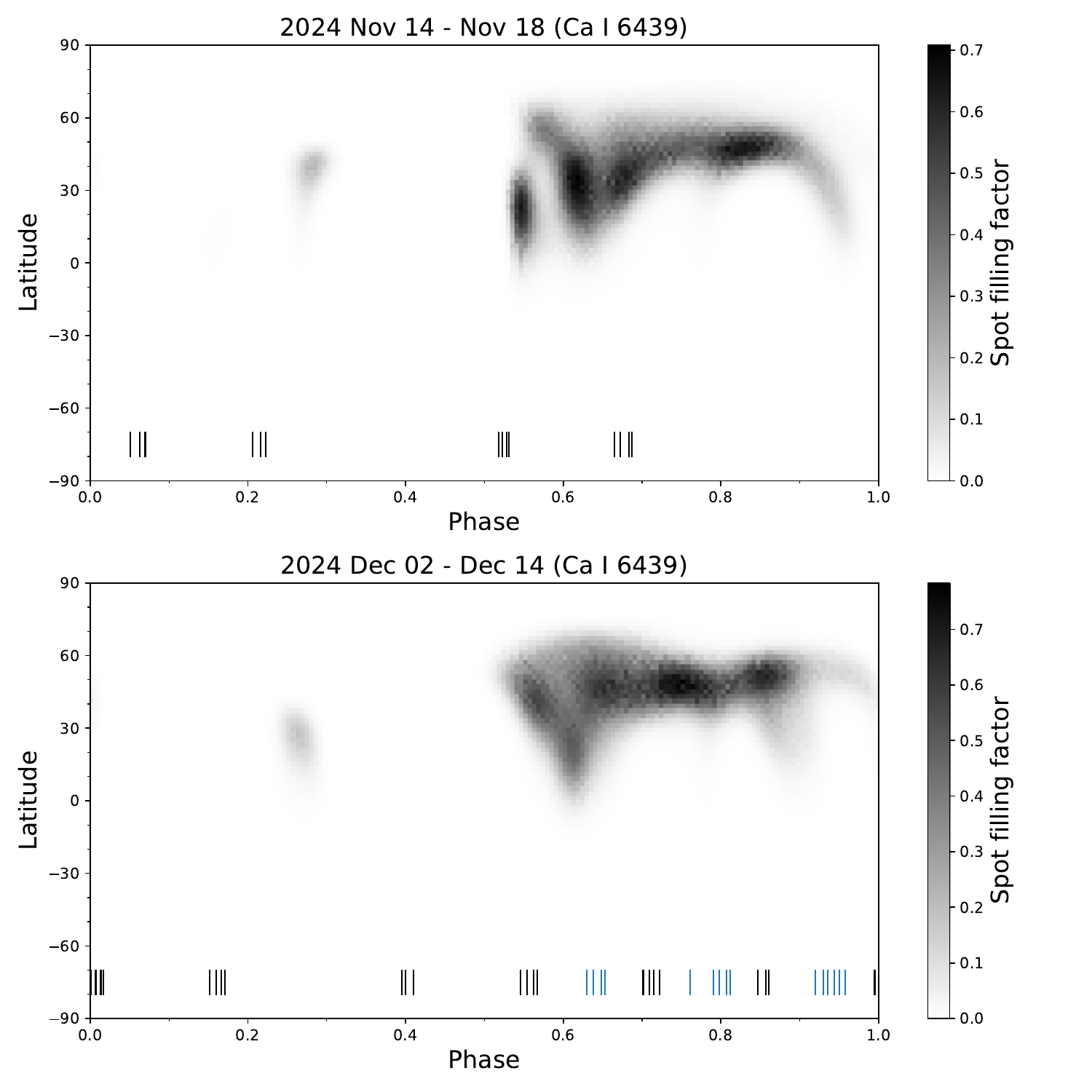}{0.45\textwidth}{}
          }
\gridline{\fig{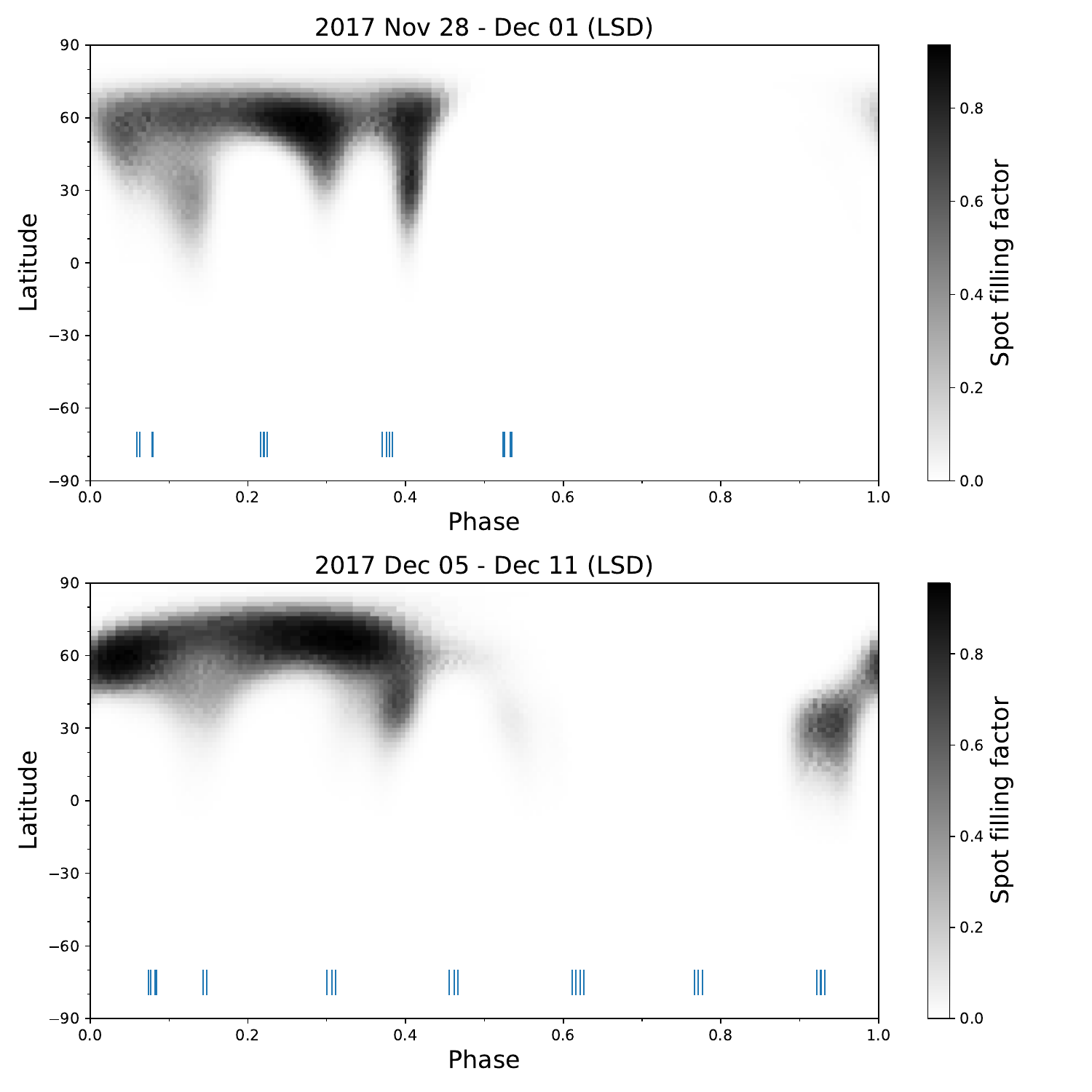}{0.45\textwidth}{}
          \fig{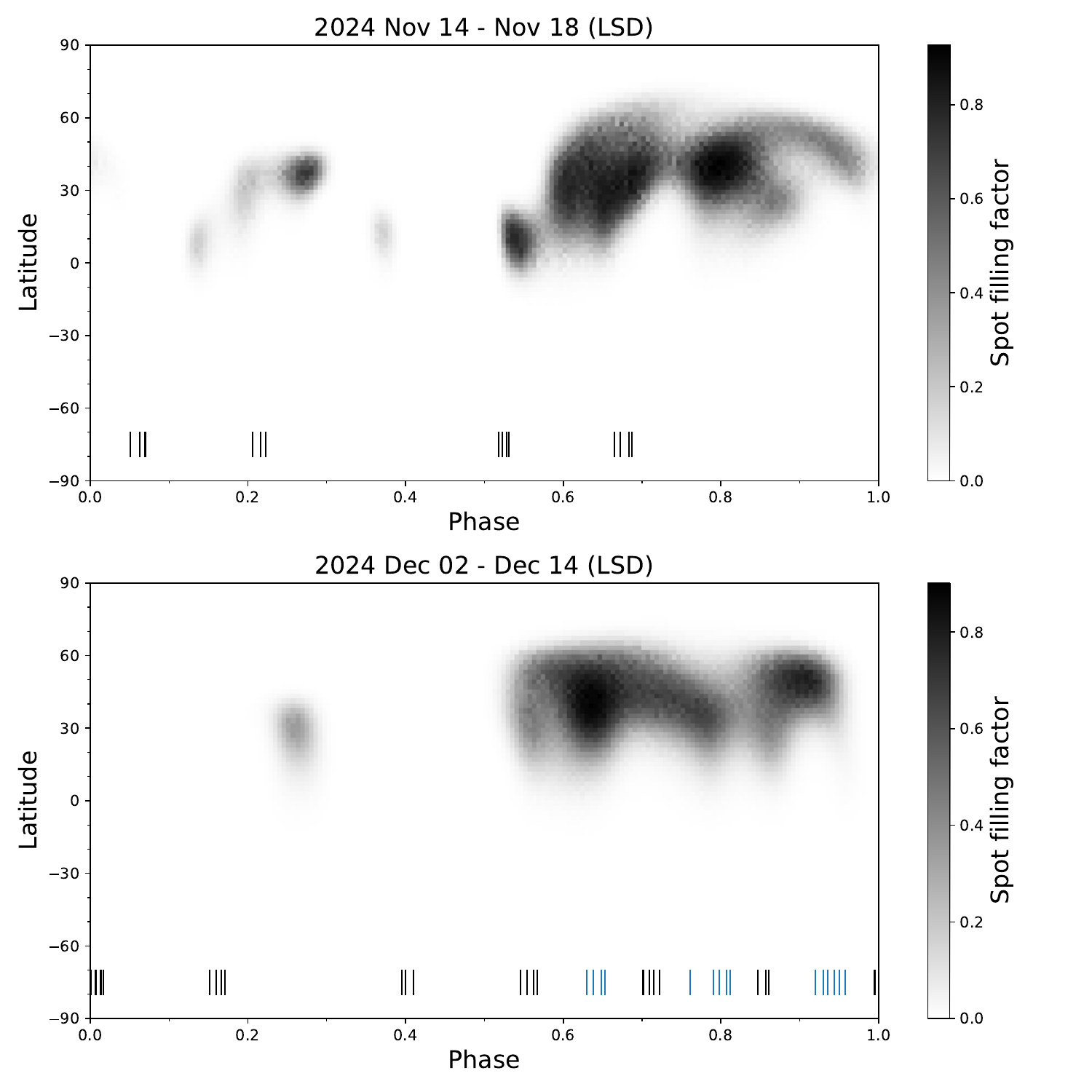}{0.45\textwidth}{}
          }
\caption{The maximum entropy regularized reconstruction of the surface images of the K0~IV component of UX~Ari for 2017 (left) and 2024 (right). The observed phases are marked as the vertical ticks in the bottom of each panel.
\label{fig:image}}
\end{figure*}

\begin{figure*}
\gridline{\fig{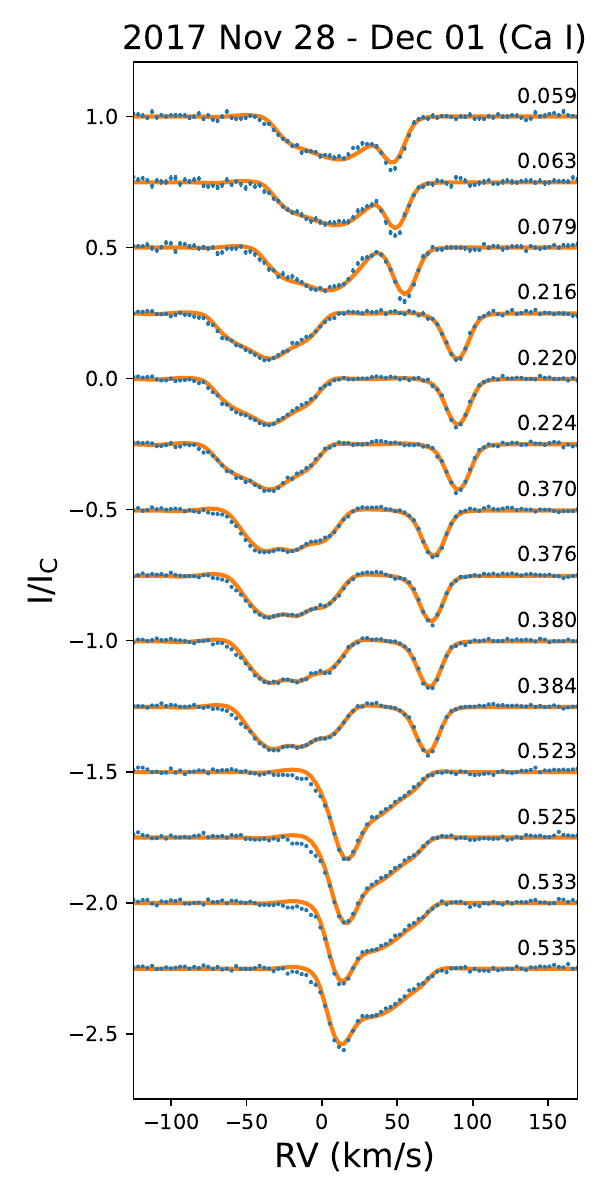}{0.23\textwidth}{}
          \fig{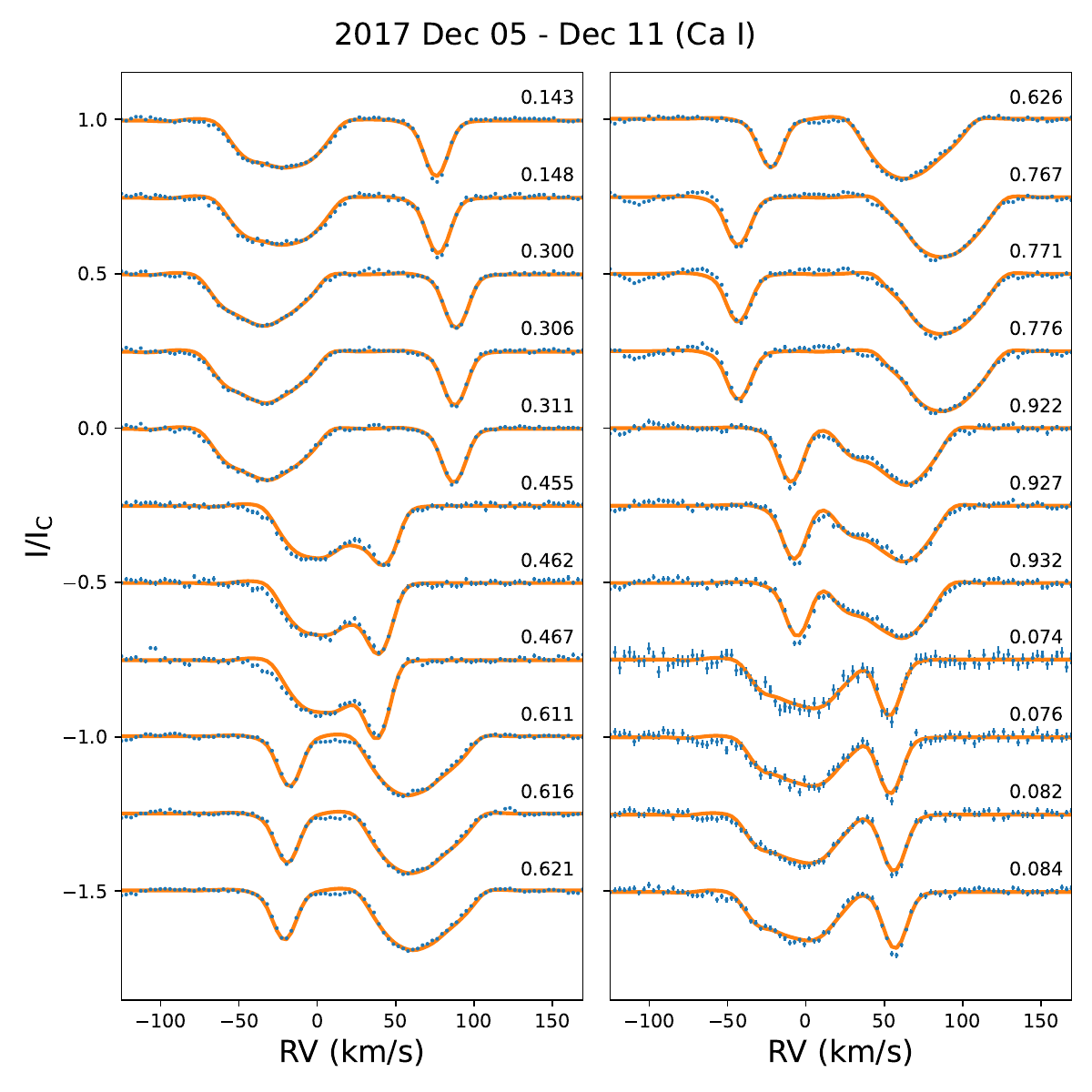}{0.46\textwidth}{}
          }
\gridline{\fig{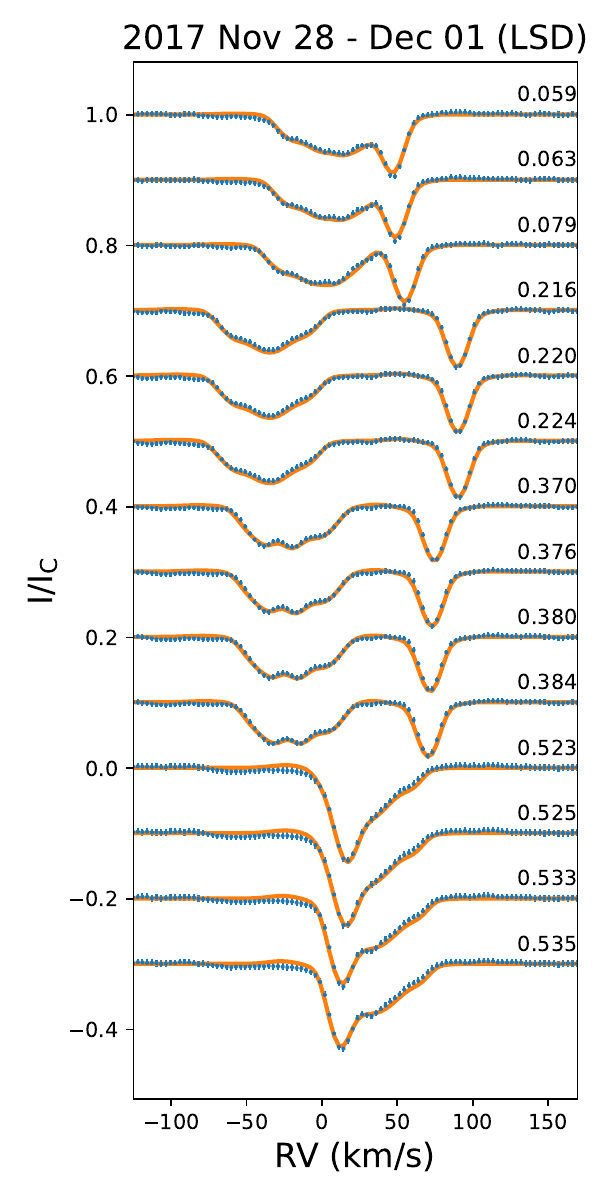}{0.23\textwidth}{}
          \fig{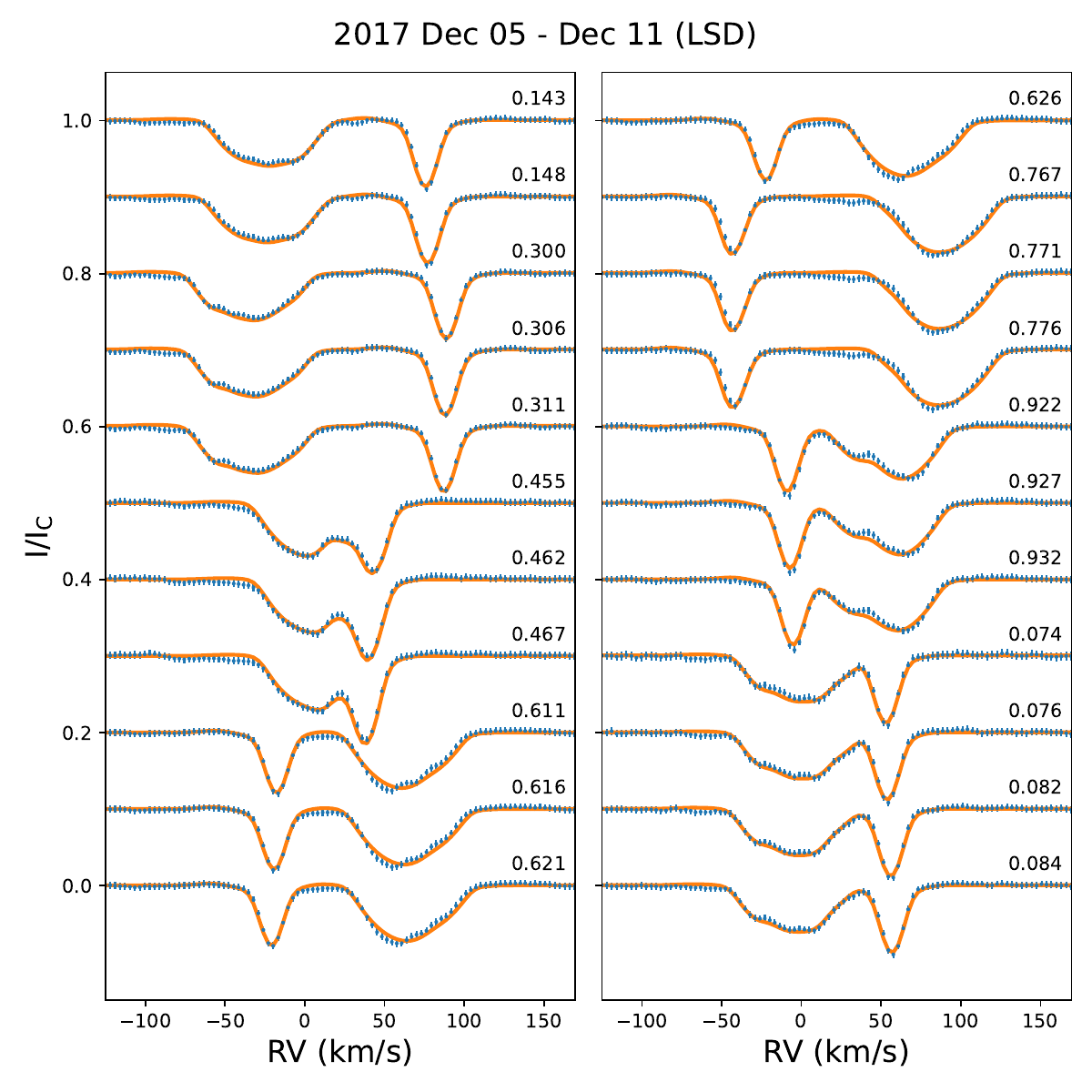}{0.46\textwidth}{}
          }
\caption{The fits to the observed Ca I 6439 \AA\ lines and the LSD profiles in 2017.
\label{fig:spec2017}}
\end{figure*}

\begin{figure*}
\gridline{\fig{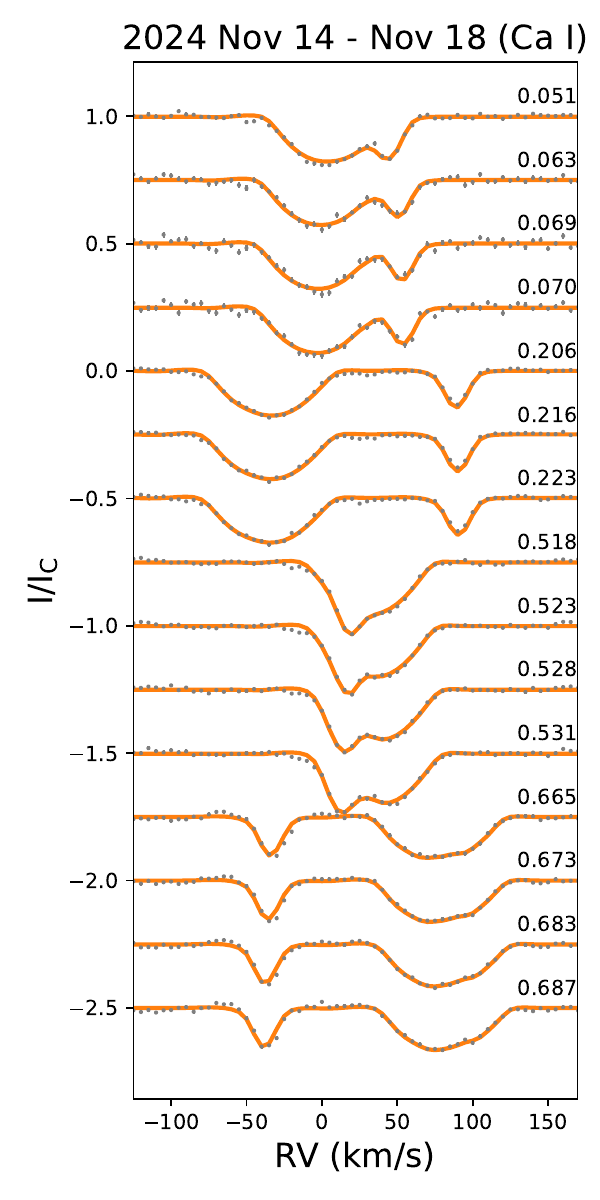}{0.23\textwidth}{}
          \fig{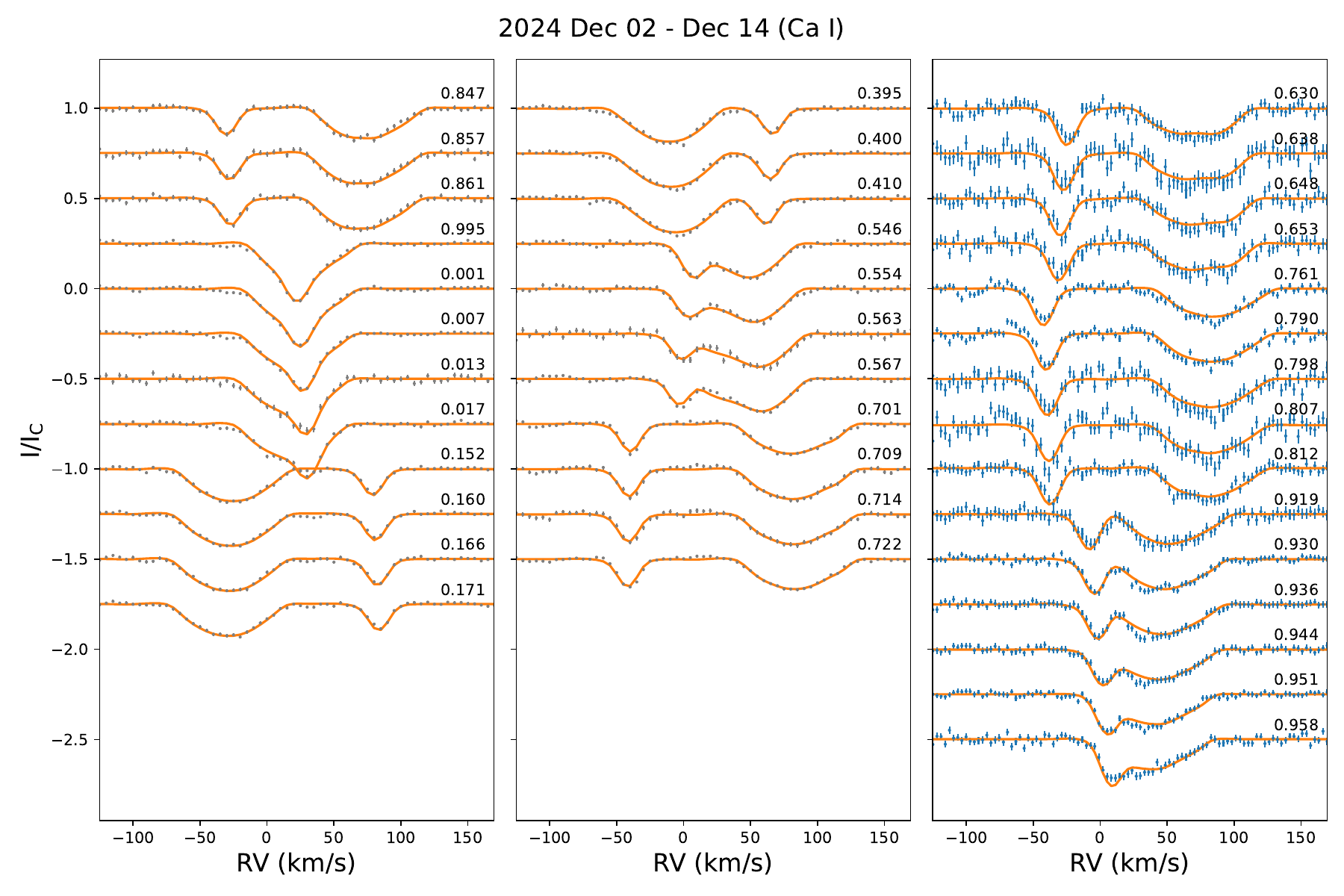}{0.69\textwidth}{}
          }
\gridline{\fig{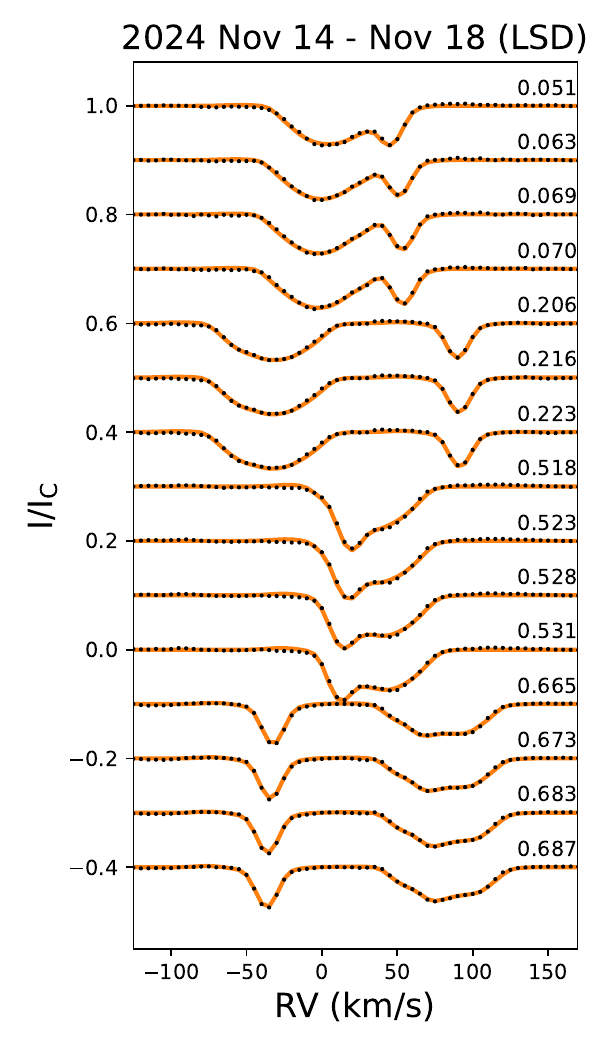}{0.23\textwidth}{}
          \fig{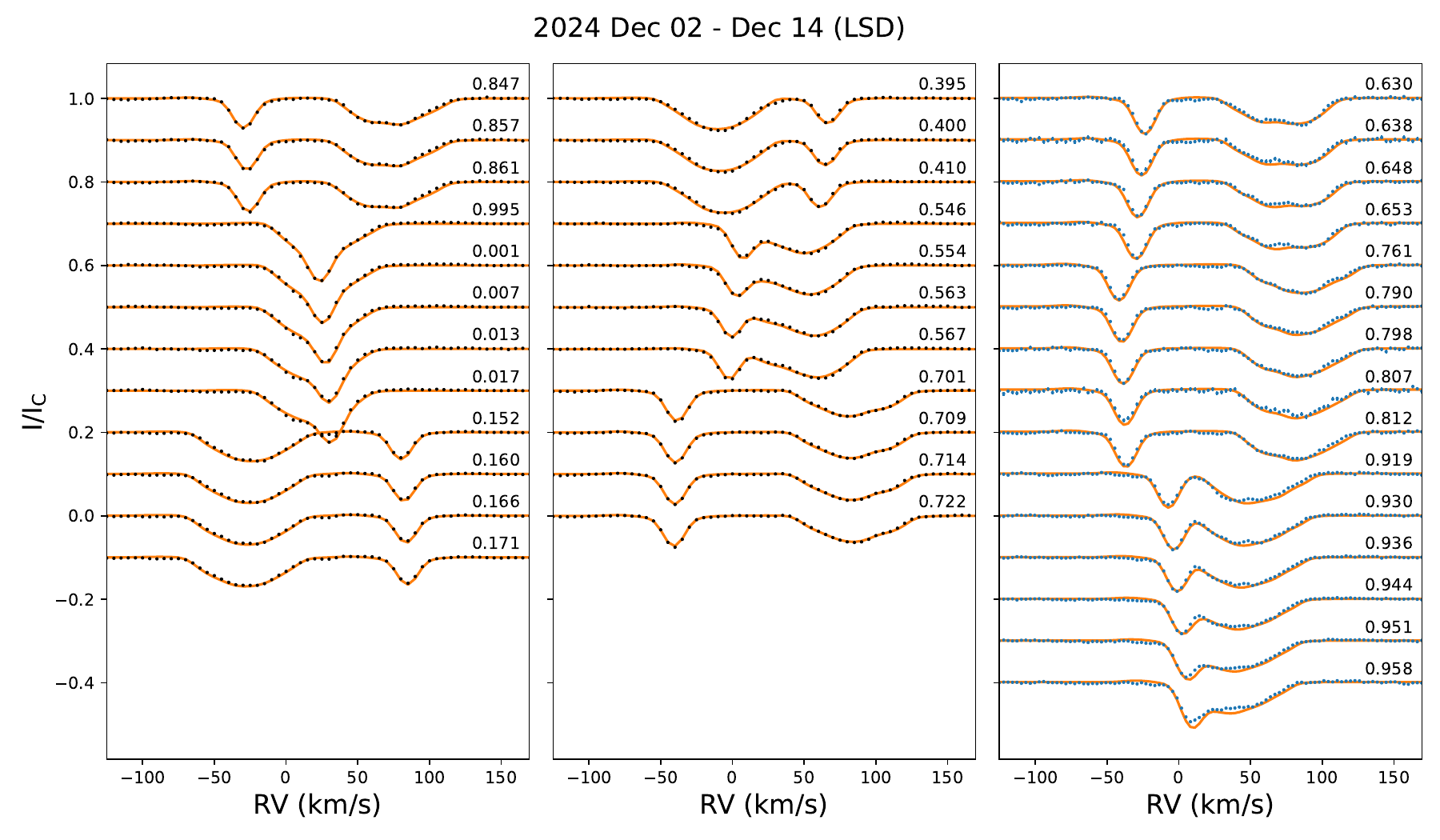}{0.69\textwidth}{}
          }
\caption{Same as Figure \ref{fig:spec2017} but for 2024.
\label{fig:spec2024}}
\end{figure*}

\begin{figure}
\centering
\includegraphics[width=0.45\textwidth]{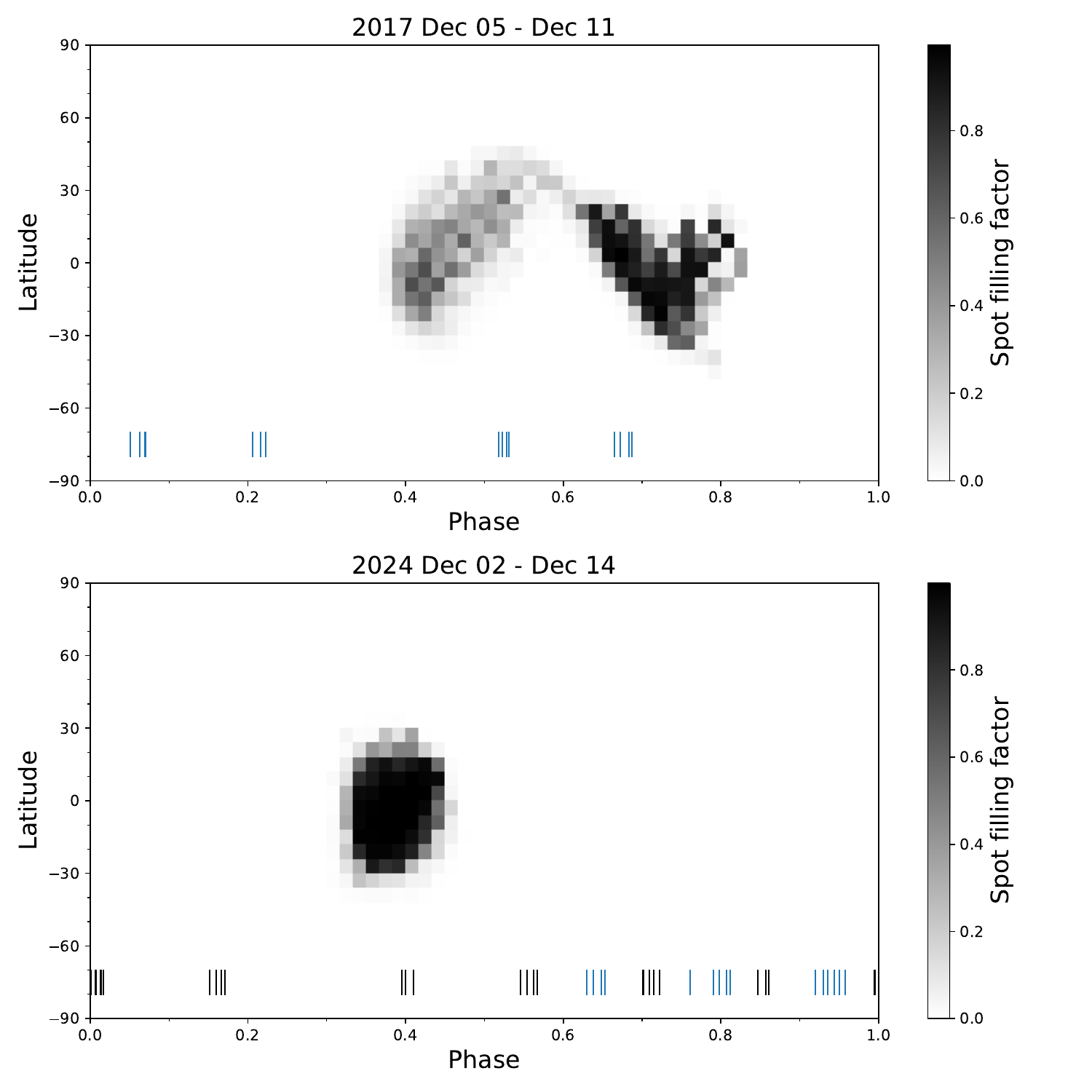}
\caption{The reconstructed starspot maps for the secondary G5~V component based on the LSD profiles.
\label{fig:sec}}
\end{figure}

For each observing run, we have divided the data into two subsets so that we can derive the surface differential rotation utilizing two consecutive surface maps. The reconstructed maps of the K0~IV component are shown in Figure \ref{fig:image}, while the fits to the observed spectral lines are presented in Figure \ref{fig:spec2017} and \ref{fig:spec2024}, where blue denotes observations from the Xinglong 2.16m telescope and black those of the Lijiang 2.4m telescope. The good agreement between the observed Ca I 6439 \AA\ lines and the modeled profiles demonstrates that the spectral line variations were resulted from rotational modulation of the starspots. With the help of the LSD profiles of high SNRs, we also show the reconstructed starspot maps for the G5~V component in Figure \ref{fig:sec}, but the slow rotation speed limits their resolution and reliability. 

The most significant feature revealed by all of the reconstructed maps is the dominant starspot group on the surface of K subgiant primary star. In 2017 November and December, the starspot group was located between phases 0 and 0.4, around latitude $60\degr$. In 2024 November and December, the starspot group was placed between phases 0.6 and 0.9, around latitude $50\degr$. There were also some appendages of the main starspot group extending to the stellar equator. However, due to the poor latitude resolution around equator for Doppler imaging, the low-latitude features are elongated vertically.

\begin{figure*}
\gridline{\fig{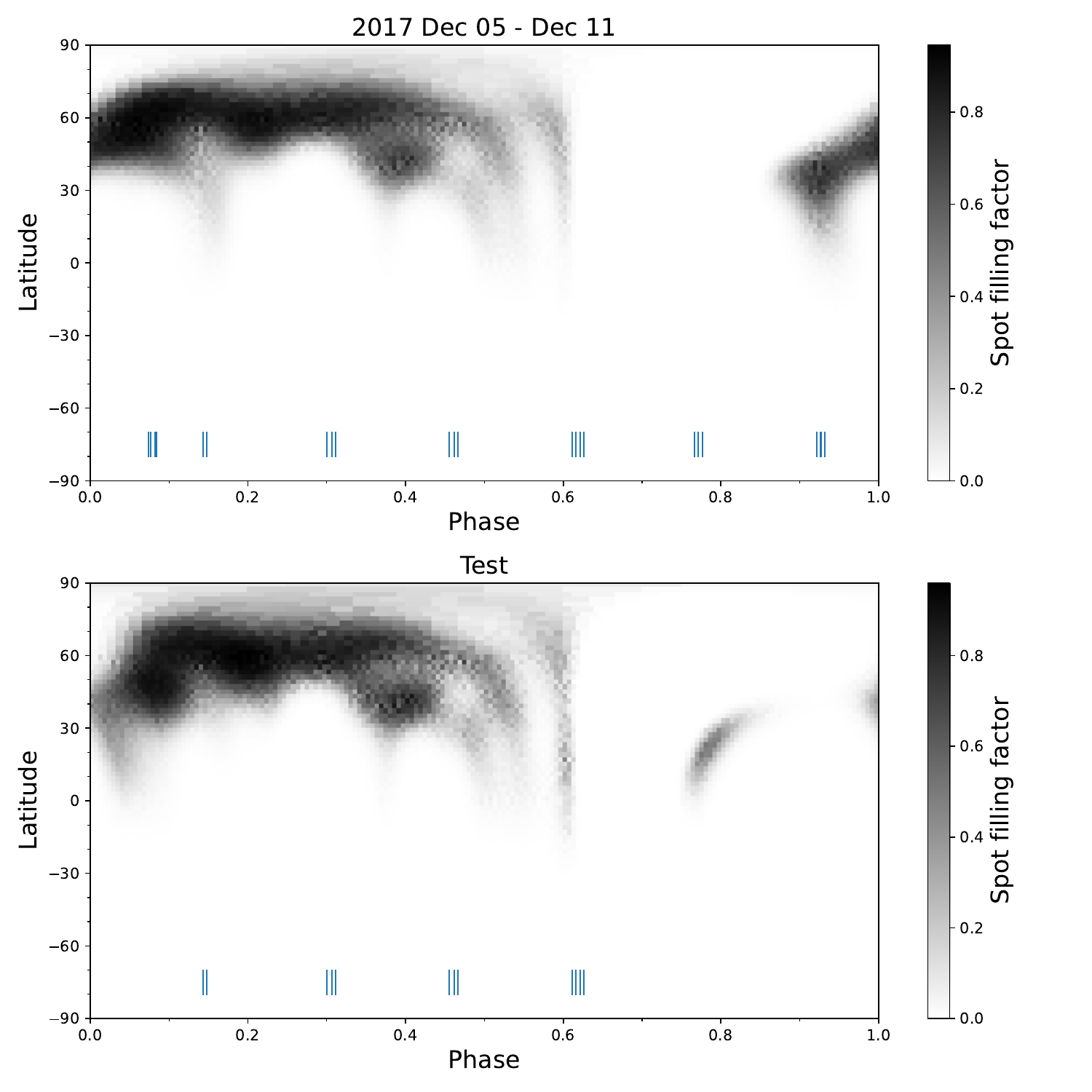}{0.45\textwidth}{}
          \fig{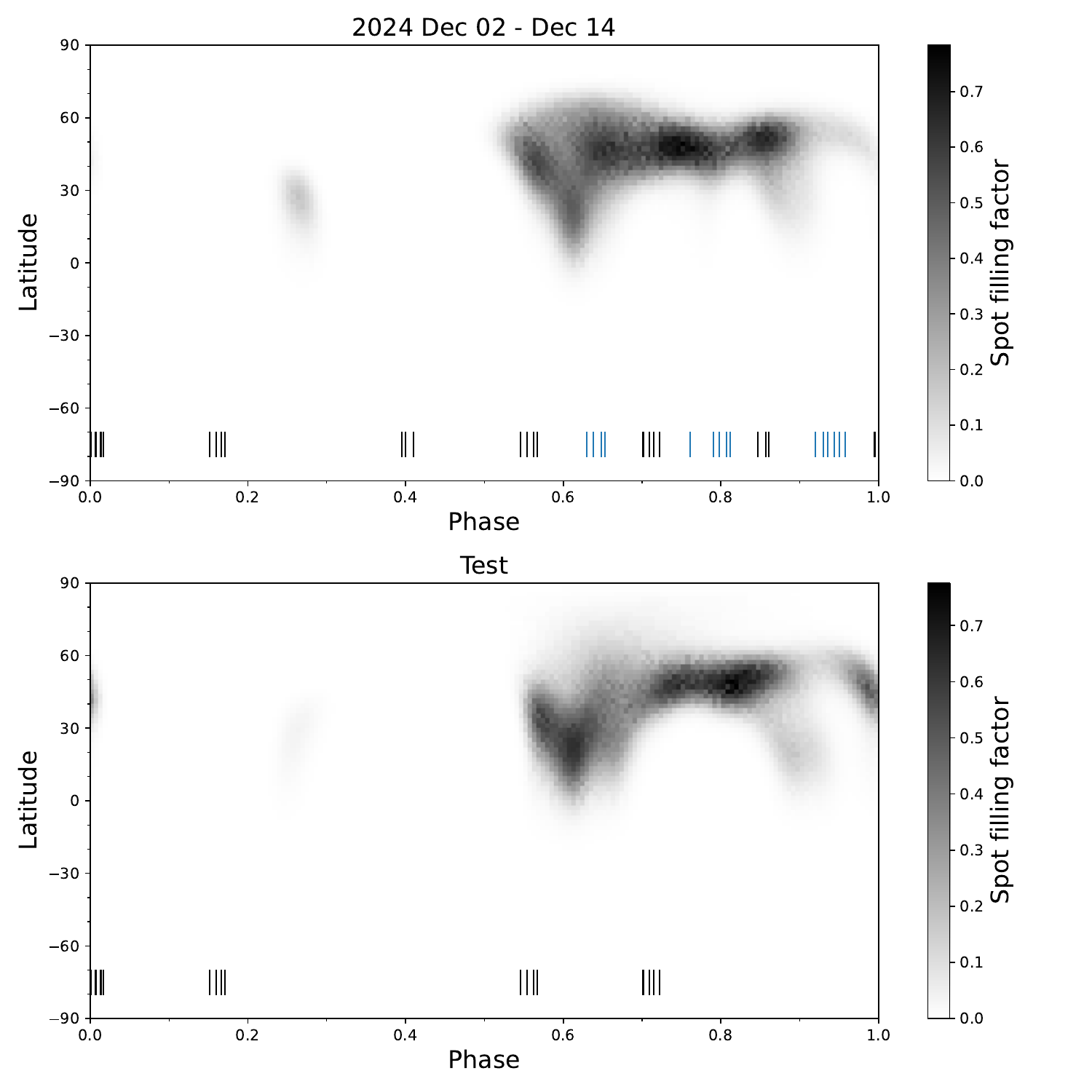}{0.45\textwidth}{}
          }
\caption{Demonstration on the effects of incomplete phase coverage on the Doppler imaging based on the Ca I 6439 \AA\ lines. Top panels are the Doppler images using the datasets with complete phase coverage in the two observing runs, and bottom ones are the corresponding images with phase gaps similar to the first subsets of two observing runs.
\label{fig:test}}
\end{figure*}

The second Doppler image of each year has a complete phase coverage, whereas the first one has a worse data sampling. The phase gap is about 0.5 in the Doppler image of 2017 November 28--December 01 and about 0.3 in the one of 2024 November 14--November 18. To check the reliability and the limitation of the Doppler images of UX~Ari derived from the datasets with poor phase coverage, we construct two test datasets with similar phase gaps by using the datasets with good phase coverage. Then we have performed Doppler imaging with the test datasets following the same procedure. The results are shown in Figure \ref{fig:test}, which demonstrate that the starspot structures are well reconstructed, especially for the 2024 dataset. Although the two subsets are only from 4 observing nights, the 2024 subset has more uniform phase coverage with smaller gaps. For the 2017 subset, the large phase gap affects the reconstruction around phases 0.9-1.0, to produce an artifact at phase 0.8 and latitude 30$\degr$, which may be due to the smearing of the real starspot.

\subsection{Differential rotation}

The comparison between Doppler images obtained at near epochs can provide clues of the latitude-dependent rotation at stellar surface \citep{donati1997a}, which is a key ingredient of the stellar magnetic dynamo process. For both 2017 and 2024, we have obtained two surface images of the primary star of UX~Ari, which allow us to estimate its surface differential rotation rate. We have derived the cross-correlation function (CCF) between the latitude bands of the two images for each year. The peak of the CCF at each latitude is measured by fitting a Gaussian profile and the error of the shift is estimated as the full width at half maximum (FWHM) of the fitted profile. To model the latitude-dependent shifts in longitude, we adopt a solar differential rotation law
\begin{equation}
\Omega (\theta) = \Omega_{eq} - \Delta \Omega \sin^{2} \theta,
\end{equation}
where $\Omega (\theta)$ is the rotational angular velocity at latitude $\theta$, $\Omega_{eq}$ is the one at the stellar equator, and $\Delta \Omega$ is the difference between angular velocities of the equator and the pole. In this definition, a positive $\Delta \Omega$ means the equator rotates faster than the pole, which is solar-like, and a negative value means an anti-solar surface differential rotation. In the fitting procedure, we only use points in the range between $35\degr$ and $75\degr$, because, on the one hand, the dominant starspots are located within this latitude range, and on the other hand, the latitude resolution at low latitudes and the longitude resolution at high latitudes are poor in Doppler imaging.

\begin{figure*}
\centering
\begin{tabular}{ccc}
\includegraphics[width=0.3\textwidth]{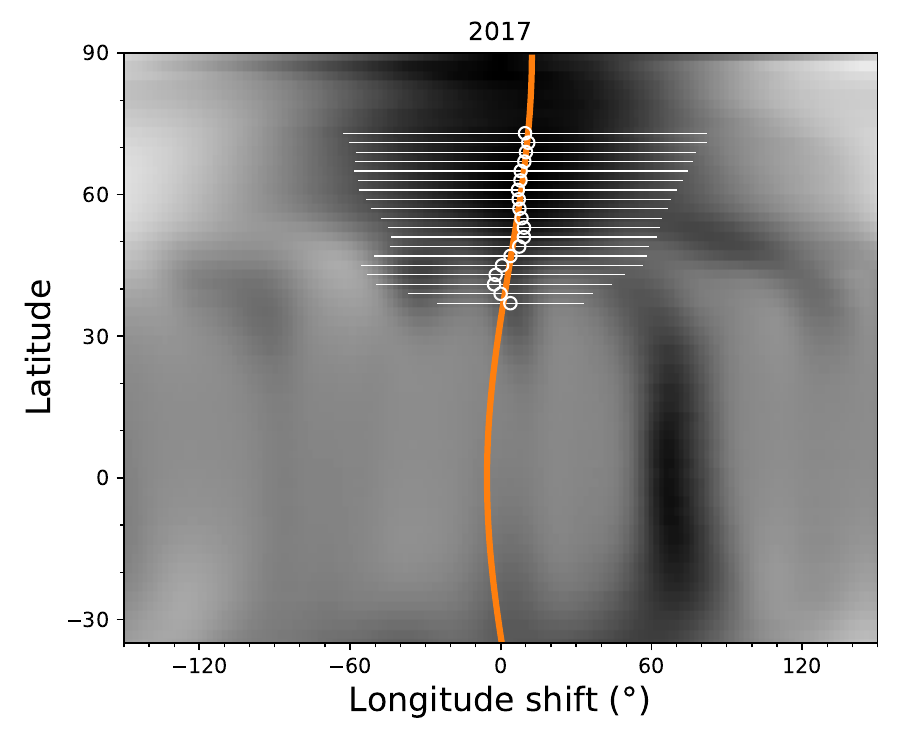}
\includegraphics[width=0.3\textwidth]{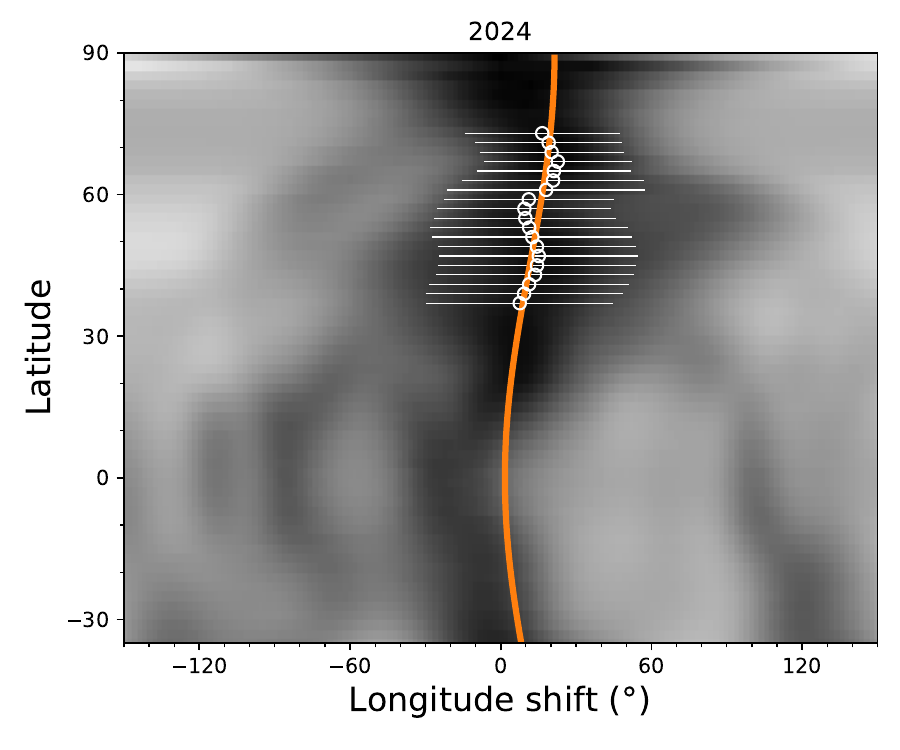}
\includegraphics[width=0.3\textwidth]{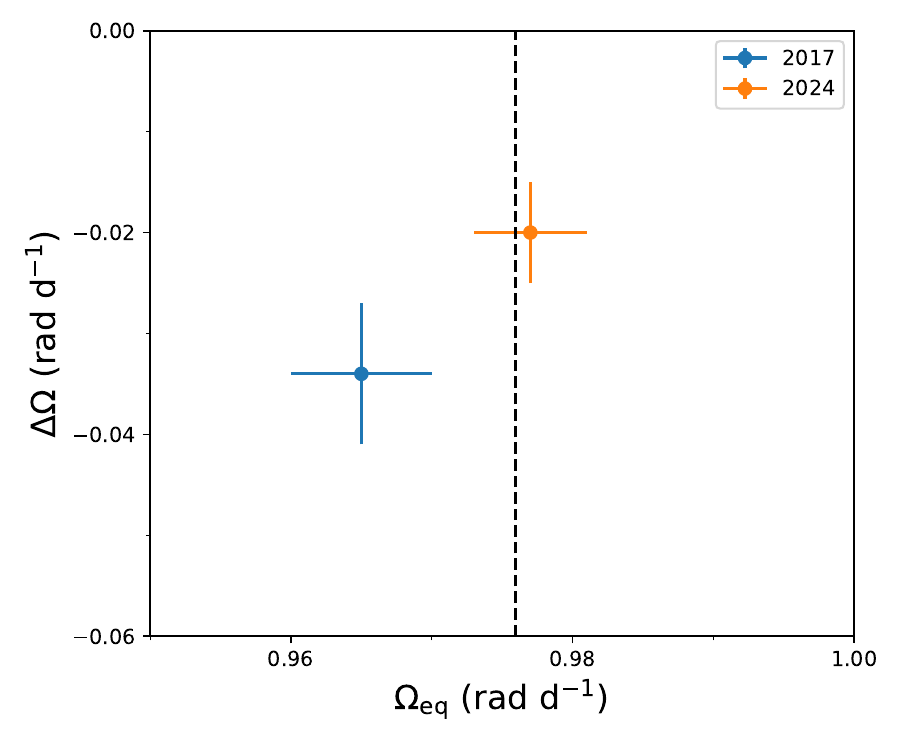}
\end{tabular}
\caption{Left two panels show the CCF maps and differential rotation fits for 2017 and 2024 Doppler images reconstructed from the LSD profiles, where the CCF peaks are indicated by white open circles and the FWHMs of the fitted Gaussian profiles are plotted as white horizontal lines. Right panel shows the estimates of differential rotation from two observing runs, where the vertical dashed line marks the synchronized angular velocity of the primary star.}
\label{fig:dr}
\end{figure*}

We display the CCF maps, the fitted latitude-dependent rotation curves, and the measured values with errors in Figure \ref{fig:dr} for 2017 and 2024. From the Doppler images of 2017 reconstructed using the LSD profiles, we have derived $\Omega_{eq} = 0.965 \pm 0.005 ~{\rm rad~d}^{-1}$ and $\Delta \Omega = -0.034 \pm 0.007~{\rm rad~d}^{-1}$. From the Doppler images of 2024 based on the LSD profiles, we have obtained $\Omega_{eq} = 0.977 \pm 0.004 ~{\rm rad~d}^{-1}$ and  $\Delta \Omega = -0.020 \pm 0.005~{\rm rad~d}^{-1}$. The difference is more likely due to the measurement errors and incomplete phase coverages rather than the evolution of differential rotation. Overall, the results indicate a weak anti-solar differential rotations on the surface of the primary star of UX~Ari.

\section{Discussion and conclusions}

We have presented new Doppler imaging of the K0 subgiant component of the RS CVn-type binary star UX~Ari, based on the spectroscopic observations in November and December of 2017 and 2024. We have derived the Doppler images from the Ca I 6439 \AA\ line and the LSD profiles with selected lines, because some of spectral lines of the component show unexplained rapid and significant variations on timescales of 1-2 hours, which seem not to be attributed to strong short-term magnetic activities like flares.

All Doppler images show a non-axisymmetric distribution of starspots on the surface of the primary star, where the dominant feature is a large mid-to-high latitude starspot group. The latitudes of these starspots were similar in 2017 and 2024, which are around 30-60$\degr$, but their locations differed by about 0.5 in the rotational phase. We do not find any significant polar starspot for the primary star of UX~Ari, in contrast to the results of \citet{vogt1991}. The starspot distribution revealed by our Doppler images is similar to that revealed by \citet{hummel2017}. The G5 main-sequence component of UX~Ari is also magnetically active. However, its relatively low rotational velocity prevents to derive more reliable surface images with good resolution.

\citet{cao2025} reported a filament eruption event on the primary star of UX~Ari on 2017 December 28, only half a month after the observations we have employed. They observed the filament eruption occurred near phase 0.7, while a huge flare were on the opposite hemisphere around phase 0.3 at a latitude of 50$\degr$, observed from 2017 December 30 to 2018 January 03. The surface maps we have reconstructed for 2017 November–December suggest that the dominant starspot structure appears to be associated with the large flare event observed just half a month later, as they were located in the same region. \citet{cao2025} also revealed that the H$\alpha$ emission of UX~Ari was at a low level on 2017 December 08, which is the epoch of our Doppler imaging. The correlation between starspots and flares has been established in many studies of different types of stars through light curve analyses \citep{araujo2023,tokuno2025}. \citet{elias1995} found that the flares on the surface of the primary star of UX~Ari took place near the starspot groups, based on both radio observations and optical photometry, indicating an association between flares and starspots in this system.

\begin{figure*}
\centering
\includegraphics[width=0.45\textwidth]{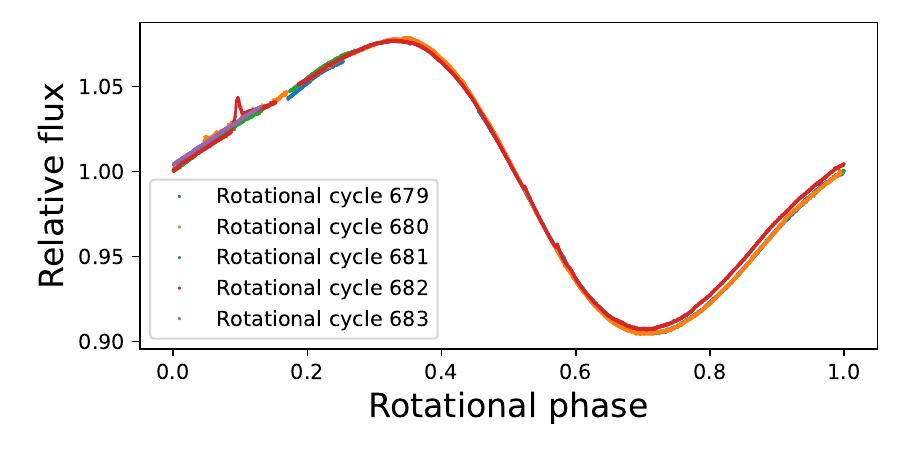}
\caption{Phase-folded TESS light curves of UX~Ari from 2024 October 27 to November 21.}
\label{fig:lcp}
\end{figure*}

The time intervals between the two Doppler images are about 9 days in 2017 and 17 days in 2024, corresponding to approximately 1.5 and 3 rotational cycles of UX~Ari primary star, respectively. The starspot distribution seems to remain relatively stable over these timescales. This stability is further supported by the TESS photometric data, as shown in Figure \ref{fig:lcp}. The shape of the light curves of UX~Ari only changes marginally during the interval from 2024 October 27 to November 21. The longitude distribution of starspots revealed by our Doppler images is consistent with the light curve in 2024 November-December, in which the light minimum occurs around phase 0.7. Our Doppler images indicate the presence of a preferred active longitude on the primary star of UX~Ari. However, given the 7-year gap between the two observing runs, we can not address any conclusions on the migration of the active longitude. \citet{cao2017} revealed that the chromosphere of UX~Ari exhibits active longitudes, which evolved from 2001 to 2004. Tidal forces in close binary systems can influence the emergence of the magnetic flux tubes and lead to the formation of the active regions at preferred longitudes \citep{holzwarth2003}. Some RS CVn-type binary systems show migration and flip-flop phenomena of active longitudes \citep{lindborg2011,kajatkari2014}.

The differential rotation plays an important role in the formation and evolution of stellar magnetic field. Through the cross-correlation method, we have derived a weak anti-solar differential rotation on the surface of the primary star of UX~Ari. The derived differential shear rates are $\Delta \Omega = -0.034 \pm 0.007~{\rm rad~d}^{-1}$ from the 2017 Doppler images and $\Delta \Omega = -0.020 \pm 0.005~{\rm rad~d}^{-1}$ from the 2024 Doppler images, which are reconstructed from the LSD profiles. The relative shear rates are $\alpha = \Delta \Omega / \Omega_{\rm eq} = -0.035 \pm 0.007$ and $-0.021 \pm 0.005$ for 2017 and 2024, respectively. The weak anti-solar shear revealed by our Doppler imaging is in good agreement with the result of \citet{vogt1991}, who found a relative differential rotation rate of $\alpha = -0.02 \pm 0.002$ by tracing individual starspots on the primary star of UX~Ari. This places the subgiant component of UX~Ari in the anti-solar differential rotation regime. The pole of the primary star laps the equator about 0.5-1 year. The rotation speed of its equator can match the synchronized rotation within the error, indicating the equator is well tidally locked to the secondary star.

While several giant stars have been reported to exhibit anti-solar differential rotation from Doppler imaging studies \citep{kovari2013,kriskovics2014}, a few subgiants in RS CVn binaries have shown this feature. \citet{harutyunyan2016} detected the anti-solar shear on the surface of the active K0 subgiant component of the RS CVn-type system HU Vir, which confirmed the measurements of \citet{strassmeier1994} and \citet{hatzes1998}. The relative differential rotation rate of UX~Ari is similar to that of HU Vir. The anti-solar differential rotation may be a result of the strong meridional flow \citep{kitchatinov2004}. Through simulations, \citet{gastine2014} demonstrated that the cool stars with large Rossby numbers exhibit anti-solar differential rotation. However, the samples are still insufficient so far, and more detections of anti-solar differential rotation are required so as to clarify its mechanism.

The phase coverage limits the precision of our estimation for the surface differential rotation on UX~Ari. Our long-term monitoring program of active stars has been ongoing for years. In the future, this will allow us not only to determine the rotation for UX~Ari with a higher accuracy but also to detect surface shear rates in a larger sample of stars.

\begin{acknowledgments}
We are grateful to the anonymous referee for the insightful comments and suggestions, which led to significant improvements of the manuscript. This work is supported by the National Natural Science Foundation of China under grant No. 12288102. The present study is also financially supported by the National Natural Science Foundation of China under grants Nos. 10373023, 10773027, U1531121, 11603068, 11903074 and 12373039, the Yunnan Fundamental Research Projects (grant Nos. 202201AT070186 and 202305AS350009), the Yunnan Revitalization Talent Support Program (Young Talent Project), International Centre of Supernovae (ICESUN), Yunnan Key Laboratory of Supernova Research (No. 202505AV340004), and the China Manned Space Program with grant No. CMS-CSST-2025-A15. We acknowledge the support of the staff of the Xinglong 2.16m and the Lijiang 2.4m telescopes. This work is partially supported by the Open Project Program of the Key Laboratory of Optical Astronomy, National Astronomical Observatories, Chinese Academy of Sciences. Funding for the 2.4m telescope has been provided by Chinese Academy of Sciences and the People's Government of Yunnan Province. This paper includes data collected with the TESS mission, obtained from the MAST data archive at the Space Telescope Science Institute (STScI). Funding for the TESS mission is provided by the NASA Explorer Program. STScI is operated by the Association of Universities for Research in Astronomy, Inc., under NASA contract NAS 5–26555. All the {\it TESS} data used in this paper can be found in MAST: \dataset[10.17909/5tda-dk56]{http://dx.doi.org/10.17909/5tda-dk56}.

\end{acknowledgments}

\bibliography{uxari}{}
\bibliographystyle{aasjournalv7}

\end{document}